\documentstyle[aps,prl,12pt,graphics,epsfig]{revtex}

\begin{document}

\title{Simulations of Pattern Formation in Vibrated Granular Media }
\author{S. Luding(*), E Cl\'{e}ment(+), J.Rajchenbach(+), J.Duran(+)}
\address{(*) Institute for Computer Applications 1,
Pfaffenwaldring 27, 70569 Stuttgart, GERMANY\\
e-mail: lui@ica1.uni-stuttgart.de\\
(+) Laboratoire d'Acoustique et d'Optique de la Mati\`{e}re Condens\'{e}e, \\
Universit\'{e} P\& M Curie -B 86, 4, Place Jussieu, 75005 Paris, FRANCE}

\date{accepted by Europhys. Lett. 26.09.1996}

\maketitle

\begin{abstract}
We present simulations of peak pattern formation in vibrated 
two-dimensional (2D) granulates
and measure the dispersion relation of the pattern 
for various frequencies, accelerations, cell sizes, and layer heights. 
We report the first quantitative data from
numerical simulations showing an interesting dependence
of the pattern wavelength on the acceleration and the system size.
Our results are related to recent experimental findings and 
theoretical predictions for gravity waves.

{\noindent Pacs: 46.10.+z, 47.20-k, 05.60+w }

\end{abstract}

The dynamical properties of non-cohesive granular media have attracted a lot
of interest in recent years \cite{jaeger96}. Vibrated granular assemblies 
show a variety of possible responses like surface fluidization 
\cite{clement91,luding94,luding94c,luding95b,warr95}, convection 
\cite{evesque89}, heaping \cite{douady89}, 
and surface waves 
\cite{fauve89,pak93,melo94,clement96,goldshtein95b,wassgren96},
all phenomena being observed in both 2D and 3D systems.
In 3D experiments on vibrated layers of sand,
Melo et al. \cite{melo94} find surface
patterns similar to the surface waves obtained by 
parametric excitation in regular
fluids, i.e. the Faraday instability \cite{faraday31}.
For a review concerning gravity waves in fluids see Ref.\ \cite{cross93} 
and refs. therein. The patterns in granular materials,
viewed from the top of the 3D cell, display regular structures such as
stripes, squares or hexagons. 
Recent experiments in a reduced 2D geometry on layers of aluminum 
beads \cite{clement96} have also
shown the formation of a peak pattern instability.
From these measurements of the parametric excitation, a dispersion 
relation was reported, analoguous to the findings of Melo et al. \cite{melo94}. 
No clear mechanism for the instability has been
given so far, but results seem to indicate a behavior specific to 
granular assemblies. At high frequencies or 
great layer heights, the wavelength saturates 
at a value independent of the excitation frequency. 
In a more recent study a scaling 
was proposed for the high frequency limit \cite{metcalf96}; 
the argument accounts for a dissipation
mechanism due to some granular viscosity but no clear
evidence for the validity of the scaling was given either.
Here, we report simulations of vibrated 2D arrays of  
polydisperse spheres and compare with experiments \cite{clement96}. 
Using material parameters close to experimentally reported
values, we observe that the instabilities appear rapidly and that 
the dispersion relation for different geometries, excitation,
material parameters, and initial conditions can be studied in detail. 

The model system consists of $N$ spheres of diameter $d_i$ 
($i$ = 1,...,$N$) randomly chosen from the interval
$[d(1-w),d(1+w)]$ with $d=1.5$~mm and $w$ $=0.1$.
Testing different values of $w$ we observed the surface waves 
even in the extreme monodisperse case $w = 0$.
We use cells of horizontal width $L$ and not limited in height.
The container moves with a vertical trajectory:
$z(t)=A \sin (\omega t)$, 
where $A$ is the amplitude and $f=\omega /(2\pi )$ the frequency
of excitation. The maximum acceleration of the bottom plate is
defined as the dimensionless quantity $\Gamma = A \omega^2 / g$, with 
the gravitational acceleration $g$.
The layer thickness is $H=N d/L$ with the 
dimensionless system width $L/d$. 

Here, an event driven (ED) method is used to simulate the dynamics of 
rigid {\it hard} spheres with no intrinsic material elasticity, i.e. 
the duration of a contact is zero ($t_c = 0$).
This choice is made for two reasons: Firstly, it is crucial 
to show that the instability presented here, is decoupled from 
a possible parametric excitation of collective elastic modes which might 
be generated in the elastic network formed by a dense packing of 
{\it soft} spheres ($t_c > 0$), i.e. the ``detachment effect'' 
\cite{luding94d}. 
Secondly, in the range of response where the typical separation
between the beads is not too small, the ED simulation scheme is quite
efficient. The method's principles are the following: under
the influence of gravity, the particles follow a parabolic trajectory, 
until an event occurs. An event is either a collision 
between two particles, or a collision
of one particle with a wall or the bottom plate. From the velocities 
just before the event, the velocities after this event are computed,
accounting for the energy loss due to
friction, and some inelasticity of the material. 
In the tangential
direction, we account for friction, using the coefficient of friction
$\mu$ and the maximum tangential restitution $\beta _0$. This 
simplified description of
tangential dissipation introduces a coupling between the linear and the 
rotational degrees of freedom \cite{luding95b} and is consistent with recent 
experimental results on colliding particles \cite{foerster94}. 
Simulations performed with different $\mu$ and $\beta_0$ 
values show the instability, even for no rotational coupling at all, 
i.e. $\mu = 0$. This means that the 
rotational degree of freedom of the particles is {\em not} crucial for the 
instability to occur. Furthermore, the patterns also occur when dissipation and 
friction at the walls is switched off, 
proving that the pattern forming instability
is not influenced by the wall's properties, as e.g. convection is. 
Switching off friction
with the bottom leads to less stable patterns in the sense, that the peaks
move in the horizontal direction more easily.

In contrast to previous ED simulations of granular 
assemblies \cite{luding94c,luding95b,lubachevsky91,duran96}, 
we implement a dissipation model which uses a velocity dependent restitution
coefficient. Such a model is qualitatively consistent with experimental 
measurements
of binary collisions, reporting a restitution coefficient approaching
unity with decreasing velocity \cite{goldsmith60,kuwabara87,brilliantov96}. 
The specific model we consider here is the limiting case of a 
visco-elastic interaction law for the contact of spherical particles.
The variation of the surface of contact during the
interparticle penetration causes a non-linear elastic force,
i.e. the Hertz model \cite{landau67}, and a non-linear dissipative force, i.e.
the Kuwabara-Kono model \cite{kuwabara87}. 
Thus we use here the coefficient of normal restitution 
$\varepsilon (u)=1-\varepsilon _0 (u/u_0)^\gamma$,
with the relative velocity in the normal direction $u$, and the power
$\gamma$ = +1/5. In order to model aluminum spheres we use 
$\varepsilon _0$ = 0.4, which leads for typical velocities of $u_0$ = 1m/s
to $\varepsilon $ = 0.6. Larger velocities yield smaller coefficients
of restitution. Other models, e.g. based on the hypothesis
of plastic deformation, lead to qualitatively similar behavior
\cite{walton86}, but so far, experiments do
not provide a discriminant conclusion about the exact velocity 
dependence \cite{foerster94,goldsmith60,walton86}. 

However, this systematic decrease of dissipation with decreasing average
relative energy is not sufficient to keep the collision frequency small
under all conditions. The divergence of collision frequency, connected to
a (possibly local) loss of relative energy is usually referred to
as the "inelastic collapse". In order to keep our simulations out of the
regime of the inelastic collapse, we also introduce a cut-off time $t_x$, 
prohibiting dissipation for a second collision within 
a time-interval shorter than $t_x$.
In order to test the sensitivity of the method to the 
exact value of this cut-off time, we have compared identical simulations 
using only different $t_x$ values $t_x$ = 0s, $10^{-6}$s, $10^{-5}$s, and 
$10^{-4}$s. Except for the largest $t_x$ values, we get 
quantitatively the same results. Note that surface waves may occur 
also in the case of traditional ED simulations 
with extreme values $t_x = 0$s and $\varepsilon = const$, i.e. $\gamma = 0$.
In general, the computational effort decreases with increasing
cut-off time, and allows simulations with rather large values
of $H$. The authors are aware that more detailed studies
are neccessary to investigate all the implications of the ED-extensions
used here; however, this problem is out of the scope of this paper.

\begin{figure}[tbp]
\psfig{figure=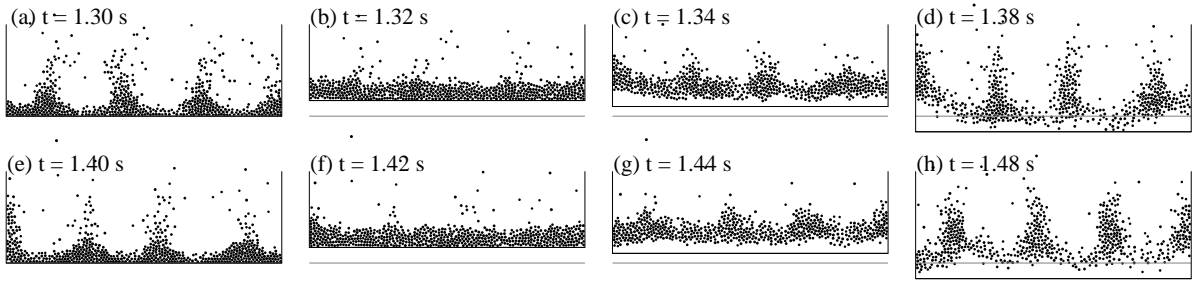,height=4.2cm,angle=0}
\caption{ Snapshots of a typical simulation over two periods
from $t$ = 1.30 s to 1.48 s.
The parameters are $N$ = 600, $L/d$ = 100, $f$ = 10 Hz, $\Gamma$ = 3.6,
$\varepsilon_0$ = 0.4, $\varepsilon_{0w}$ = 0.2, $\mu$ = 0.2,
and $\beta_0$ = 0. The dashed line indicates $z = 0$.}
\label{fig:fig1}
\end{figure}

In Fig.\ \ref{fig:fig1} snapshots of a typical simulation with $N$ = 600
particles, in a box of width $L/d$ = 100, vibrated with $f$ = 10 Hz 
and the acceleration $\Gamma$ = 3.6 are plotted. 
The parameters of dissipation and friction are here 
$\varepsilon_0$ = 0.4, $\varepsilon_{0w}$ = 0.2, 
$\mu = \mu_w$ = 0.2, and $\beta _0 = \beta _{0w}$ = 0.0, where
the index $w$ indicates the particle-wall interaction parameters,
corresponding to strong
particle-particle and weaker particle-wall dissipation. 
In Fig.\ \ref{fig:fig1}, we
present a time-series ranging from $t = 1.30$s to $1.48$s. 
We observe, like in the experiment, a
parametric response of the layer with a period $2T=2/f$. When the bottom 
moves up, see
(a) and (e), the array is compressed and the peaks vanish,
see (b) and (f). The array separates from the bottom plate after
the latter accelerates downwards, 
see (c) and (g), and the peaks grow, see (d) and (h), until
the array hits the bottom again. Note that 
the position of the peaks is interchanged with the position of 
the valleys from one period to the next. 
An arch-like structure below the array, just before the collision with the 
bottom plate, is also visible, see (h). This behavior is in agreement 
with the experimental findings of Cl\'ement et al. \cite{clement96}.

To perform systematic quantitative measurements on the
wavelength of the observed pattern, we monitor the 
behavior of the horizontal particle-particle correlation function:
$C_{x,x}(x)={\frac 1{{(L-x)N^2}}}\Sigma
_{i=1}^N\Sigma _{j=1}^N~\delta \left[ {x -|x_i-x_j|}\right]$,
with the delta function $\delta [x]=1$ for $x = 0$ and $\delta [x]=0$ elsewhere. 
If the instability is present, this function displays some 
modulation in $x$ with a first maximum which we indentify 
with the wavelength $L_x(t)$ of the surface waves. 
The modulation is strongest just before the collision 
of the granular layer with the bottom plate. 
We trace $L_x(t)$ over 20 to 50 periods and get
$L_x$, the averaged wavelength.

\begin{figure}[tbp]
\psfig{figure=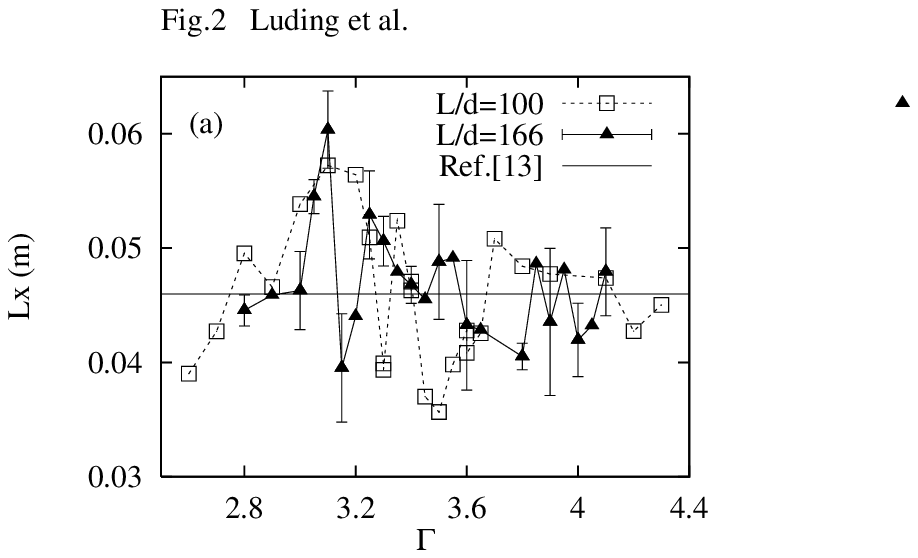,height=5.2cm,clip=}
\psfig{figure=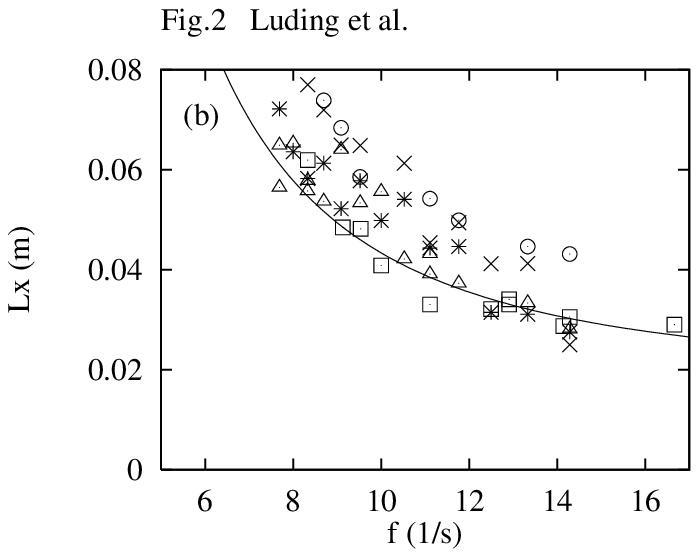,height=5.2cm,clip=}
\psfig{figure=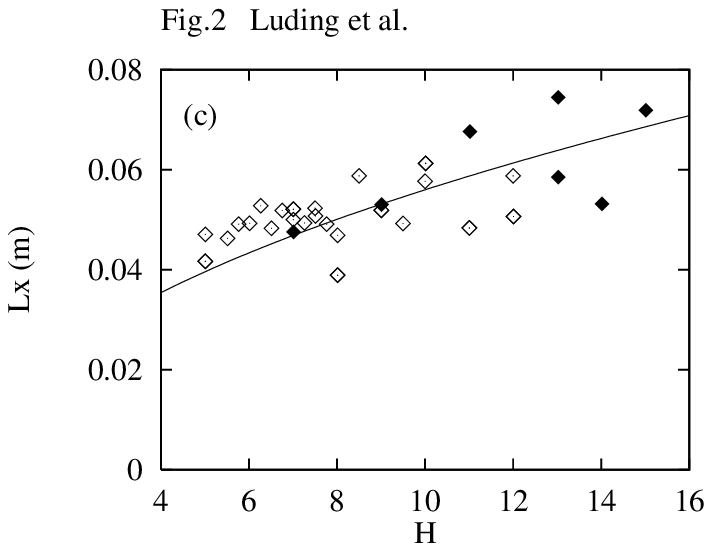,height=5.2cm,clip=}
\psfig{figure=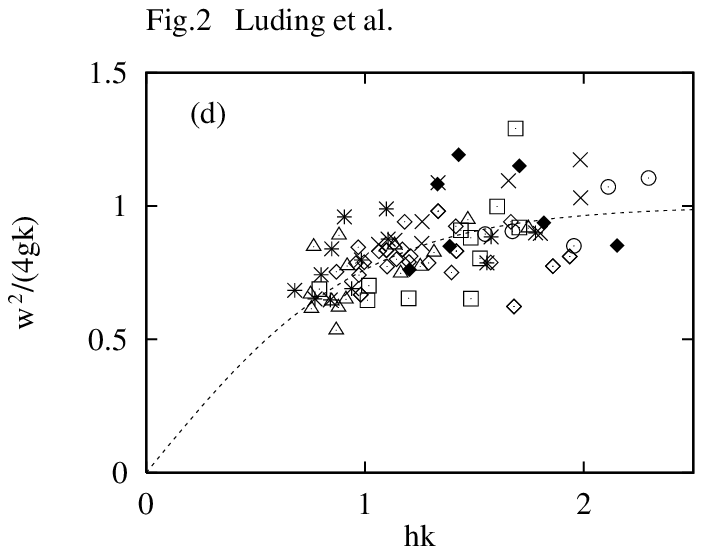,height=5.2cm,clip=}
\caption{
(a) $L_x$ as a function of $\Gamma$ for $H = 6$, $f = 10$Hz,
and $L/d$ = 100 (squares), or $L/d$ = 166 (solid triangles).
The error-bars denote the standard deviation of selected averages
and the horizontal line
is the empirical fit of Ref.\ \protect \cite{clement96} in
Figs. \ \protect\ref{fig:fig2}(a), (b), and (c).
the error-bars denote the standard deviation of selected averages.
(b) $L_x$ as a function of $f$ for $H = 6$, and $\Gamma = 3.6$,
$L/d$ = 100 (squares), $\Gamma = 3.6$, $L/d$ = 200 (triangles),
$\Gamma = 3.2$, $L/d$ = 200 (stars), and $\Gamma = 3.6$,
$L/d$ = 200, $H = 10$ (x), and $H = 14$ (circles) . The solid curve
is the empirical fit of Ref.\ \protect \cite{clement96} for $H$ = 6.
(c) $L_x$ as a function of $H$ for $f = 10$Hz, $\Gamma = 3.6$, and
$L/d$ = 100 (diamonds) or $L/d$ = 200 (solid diamonds).
(d) Collection of all simulations from Figs.\ \protect\ref{fig:fig2}(b)
and (c), in dimensionless scale. The dotted line is the dispersion relation
expected for gravity waves.
}
\label{fig:fig2}
\end{figure}

Cl\'ement et al. \cite{clement96} observe from experiment 
little influence of the acceleration $\Gamma$ 
on the dispersion relation of the waves. This was
found for aluminum beads in a range of accelerations where 
the frequency of layer-bottom
collisions roughly equals the frequency of the bottom-plate, i.e.
$\Gamma \le 4.5$.
Varying the cell width from 100 to 200 bead diameters, no influence 
was reported either. Thus an empirical dispersion relation was proposed 
to fit all the data ranging from $H = 3$ to $H = 9$ layers, i.e.
$\lambda =\sqrt{H}(\lambda ^{*}(d)+g^{*}/f^2)$,  
with $\lambda ^{*}(d)$ = 7.2 mm, and $g^{*}$ = 1.05 m/s$^2$.
In contrast, the simulations show that the wavelength does
depend on $\Gamma$. In Fig.\ \ref{fig:fig2}a, we plot the wavelength 
$L_x$ of a system with $H = 6$ at constant frequency 
$f = 10$Hz and for accelerations in the range
$2.6 \le \Gamma \le 4.3$. 

The solid flat line is the empirical estimate 
of Ref.\cite{clement96}. For $L/d = 100$ (squares) a modulation of the 
wavelength is clearly evidenced and also for $L$ increased by a non-integer 
factor, i.e. $L/d = 166$ (solid triangles), this modulation exists, 
however with a different structure. We give some typical standard 
deviations of the averages, to show that the structure is not only due
to noise. Since $L_x$ is averaged over many periods, not much 
quantitative can be said about $L_x(t)$. For the simulations with a large 
standard deviation we mostly observe a significant 
periodic change of $L_x(t)$ from period to period between two 
values.
In Fig.\ \ref{fig:fig2}b we plot $L_x$ versus $f$ for various simulations
with $H = 6, 10, 14$, $\Gamma = 3.2, 3.6$ and $L/d = 100, 200$.
Comparing the results with the empirical fit of Ref.\ \cite{clement96} 
for $H = 6$, we observe qualitative agreement with the corresponding
simulations. 
Now, we vary the height of the layer from $H$ = 5 to $H$ = 15. The
acceleration is kept constant at $\Gamma$ = 3.6 and the system width is 
$L/d$ = 100 (diamonds) or $L/d$ = 200 (solid diamonds). 
In Fig.\ \ref{fig:fig2}c, the wavelength is plotted 
as a function of $H$. We observe an increase of $L_x$ with $H$
but we cannot extract a functional behavior from our data due to the
strong fluctuations. 
In Fig.\ \ref{fig:fig2}d, the dispersion relation for fluid gravity waves 
is tested \cite{goldshtein95b}. We plot $\omega^2/(4 g k)$ as a function 
of $h k$ with $h = \sqrt{3} d H / 2$, 
and the wavenumber $k = 2 \pi / L_x$, for the simulations of
Figs.\ \ref{fig:fig2}b and c. The dotted line is the expected 
dispersion relation for gravity waves
${{\omega^2} / (4 g k)} = {\tanh ( h k ) }$.
Even when the data seem to gather near the line, the
fluctuations are too strong to allow for a conclusive 
statement on the dispersion relation.
 
In this letter, we present simulations of vibrated 2D layers of grains, 
using an algorithm  
based on an event driven procedure with a restitution coefficient depending
on the impact velocity and a collision frequency dependent dissipation
threshold. 
The simulation procedure keeps us, by algorithmic construction,
out of the regime where multiparticle effects become dominant. 
However, we verify that the patterns also occur  
with a standard interaction model. The present procedure is designed
to explore more efficiently larger domains of parameter space, i.e.
the divergence of the collision frequency 
(the ``inelastic collapse'') can be
avoided. The reported patterns are in qualitative agreement with the 
experimental findings.
We observe standing peak patterns at the layer 
top and - depending on different parameters - an 
arch structure forming at the bottom. The patterns 
oscillate with twice the bottom plate period. 
The pattern wavelength was systematically extracted using the horizontal 
density correlation function. Like in the experiments, we evidence a regime 
where the wavelength decreases when the frequency increases and an almost
constant wavelength in the limit of large frequencies. 
We observe a modulation of the wavelength 
around the experimental empirical determinations,  
which is triggered by a resonant effect 
between the box size and the bottom plate acceleration. 
This effect was not reported in previous experiments in large cells but
something analogous was reported for small cells \cite{goldshtein95b}. 
For layer heights between H = 5 and H = 15, we measure a weak increase of the 
wavelength compatible with previous experimental 
determinations but the data does not allow for a definite 
quantitative conclusion. 

Finally, we remark that the picture of a completely inelastic
block - often used to describe a dissipative thick granular layer -
is not valid if the system succeeds to choose a state, 
i.e. standing waves, in which energy is not totally dissipated
during the contact with the bottom. 
The aim of fully understanding the instability presented here
is to unravel the physics governing the modes of transport of 
mass, momentum and energy in a vibrated granular material.  
An open and challenging question is to extract from these parametric
excitation studies what might 
be specific to granular assemblies and what can be understood in the general
framework of hydrodynamic instabilities.
The qualitative convergence we find here between experimental results 
and numerical computations is encouraging and calls for more detailed 
studies on both sides. 

We thank Francisco Melo 
for interesting discussions. L.A.O.M.C. is the
U.R.A. 800 of the C.N.R.S. We acknowledge the support of the EU program
"Human Capital and Mobility" and of the PROCOPE/APAPE scientific collaboration
program. S.L. acknowledges the support of the DFG, SFB 382 (A6).


\end{document}